\documentclass{article}
\newcommand{\sa}[2]{\begin{array}{c}#1\\#2\end{array}}
\newcommand{\qa}[4]{\begin{array}{c}#1\\#2\\#3\\#4\end{array}}
\newcommand{\IR}{\hbox{\rm I\kern-.2em\hbox{\rm R}}}
\usepackage{graphics}
\usepackage{amssymb}
\usepackage{amsbsy}
\begin{document}
\title{\LARGE Periodic orbit theory of\\ two coupled Tchebyscheff maps}
\author{C. P. Dettmann$^1$\thanks{Carl.Dettmann@bris.ac.uk} 
\quad and  D. Lippolis$^{1,2}$\thanks{Future address: Department of Physics, Georgia Institute of Technology, Atlanta GA 30332, USA}
\\ \small 1. Department of Mathematics, University of Bristol,\\\small
University Walk, Bristol, BS8 1TW, United Kingdom\\
\small 2. Dipartimento di Fisica, Universit\'a di Bologna,\\\small
viale C. Berti Pichat 6/2, Bologna 40127, Italy}
\maketitle
\begin{abstract} Coupled map lattices have been widely  used as models in
several fields of
physics, such as chaotic strings, turbulence, and phase transitions, as
well as in other
disciplines, such as biology (ecology, evolution) and information processing.
This paper investigates properties of periodic orbits in two
 coupled Tchebyscheff maps. Then zeta function cycle expansions are used to
compute dynamical averages appearing in Beck's theory of chaotic strings.
The results show close agreement with direct simulation for most values of the 
coupling parameter, and yield information about the system complementary to
that of direct simulation.
\end{abstract}
\section{Introduction}
Coupled map lattices (CML) are dynamical systems with discrete space and time and continuous state variables. Their most common form in applications 
is the diffusively coupled model    
\begin{equation} 
\label{general} 
\Phi_{n+1}^{(i)}=
(1-a)f(\Phi_n^{(i)})+(a/2)[g(\Phi_n^{(i+1)})+g(\Phi_n^{(i-1)})]
\end{equation} where $n$ is a 
discrete time step, $i$ is a lattice point with a 
periodic 
boundary condition, $f(\Phi)$ the local map, $g(\Phi)$ the coupling and
$a$ a continuous parameter. They prove particularly useful in 
the investigation of chaotic dynamics of spatially extended systems
(spatiotemporal chaos), but have also been employed in chemical,
biological and engineering 
modelling, as well as pattern formation.  The discrete structure in
space and time can either be the approximation of a less tractable
continuous system (as in turbulence), or may arise naturally in
applications (as in population dynamics of discrete colonies of
organisms).  Refs.~\cite{chaos,K93,physd,turbulence} give a good overview of
early work to 1998 including the above applications;
recent developments include advances in the mathematics of
the relevant Banach spaces and transfer operators~\cite{B01,B02a} as
well as ever widening applications; examples include
ecology~\cite{S02,L03,T03a}, neurons~\cite{D00,A02},
traffic flow~\cite{N01} cryptography~\cite{T03b,W03} and field
theory~\cite{Beck,Beck2}.

From the point of view of understanding the dynamics of
coupled chaotic maps, the most common CML studied are where the
local dynamics $f$ is quadratic (the ``logistic'' map); this class
of one dimensional maps is now reasonably well understood.  The
simplest example of a quadratic map is the second degree Tchebyscheff
polynomial $f(\Phi)=T_2(\Phi)=2\Phi^2-1$ which is well known to be
conjugated to a piecewise linear map with complete binary symbolic
dynamics, and hence exactly solvable.

In this paper we consider diffusively coupled Tchebyscheff maps (see
Sec. 2.2 for details) of degree 2 and 3.  These have been previously
used by Beck \cite{Beck,Beck2} in some interesting but unconfirmed
ideas concerning
dynamical generation of noise for stochastic quantisation leading to
predictions of parameters in the standard model of particle physics.
One of the authors \cite{Dettmann} subsequently showed that four of Beck's six
CML models exhibit stable synchronised time periodic states at arbitrarily
small values of the coupling parameter $a$, albeit with small basins
of attraction.  The motivation for considering these models derives
not only from possible applications, but also theoretical simplicity in
that the uncoupled state is fully chaotic and exactly solvable, the
small coupling limit has interesting structure, and the equations (being
low degree polynomials) allow many exact calculations.

Here we apply classical periodic orbit theory \cite{AAC,webbook1} to a
lattice of two
coupled Tchebyscheff maps.  This theory allows very precise computations of
long time properties (averages, Lyapunov exponents, dimensions, decay of
correlations, etc.) of chaotic systems, but has been mostly restricted
to low dimensional systems in the past.  This paper contains the first
steps towards developing a high dimensional periodic orbit theory, making
use of the properties of Tchebyscheff maps to locate a large number of
periodic orbits.  If the Beck theory turns out to be true,
the insight gained regarding these CML will be invaluable in developing
periodic orbit based computation methods which are potentially far more
precise than direct simulation, and which are required in particle physics.
More generally, a periodic orbit theory of CML will allow a far greater
precision and insight into the many spatiotemporal systems discussed at the
beginning than is currently available using direct methods.

In order to achieve our immediate goal of computing Beck's observables
on a two site lattice, we first need to look for periodic orbits
of~(\ref{main}) by using appropriate numerical methods 
and study some of their properties, as described in 
the following sections. The locations, the stability and the number
of periodic orbits of such maps depend on the value of 
$a$ and current results do not allow prediction of these changes in detail.
As we discuss in section 2, unexpected phenomena take place as soon as
the coupling 
parameter $a$ is increased from 0, for example a few periodic points
suddenly move off the region $\Phi^i\in[-1,1]$, which we consider for our 
analysis. As $a$ is increased from, periodic points drift on the
domain $\Phi^i\in[-1,1]$, moving into the complex plane and then
reappearing on the real axis as $a$ is increased further.
The stability of periodic orbits is also heavily affected by
the value of the coupling and stable and/or almost stable orbits 
appear for certain $a$'s. Such behaviour has important consequences for
the convergence of cycle expansions 
\cite{webbook1}, as discussed in section 3. 
In fact both the rate and the kind of convergence
of any average we compute on the lattice strongly depends on the
value of the coupling parameter. 
Still in the same section we address the applications of
equations~(\ref{main}) to the theory of Beck. As pointed out 
by one of the authors \cite{Dettmann}, the 
system~(\ref{main}) does not show ergodicity for all values of $a$,
therefore direct simulation is not always guaranteed to be reliable in 
computing averages over the lattice. That is one reason why we investigate the
efficiency of averaging over the periodic points. When it is possible 
to achieve a significant precision for the zeta function (see sec. 3.1),
we have computed the average of some string potentials defined in sec. 3.2, 
and then compared the results with direct simulation on the same lattice.
Conclusions and outlook are presented in section 4.     

\section{Periodic solutions}
\subsection{Finding periodic orbits}
The first problem is to find periodic solutions to the 
map~(\ref{main}) up to the highest
possible period. A number of numerical methods can be employed for that
purpose.  In our case the
quickest one seems to be the inverse iteration method, as follows:
Let $x'=f(x)$ be a map whose cycles are
all expanding\footnote{they have no marginal or contracting directions},
then it follows that
its inverse $x'=f^{-1}(x)$ has only stable periodic orbits, which
can easily be found by
iteration. A different symbol sequence (where symbols are associated
with branches of the multivalued inverse)  is to be used for every orbit 
\cite{webbook1}.
Unfortunately this
method cannot be applied a priori, unless we are certain that the map
has only expanding cycles.
It turns out that the system~(\ref{main}) has got a few periodic
orbits whose Jacobian has one
stable and one unstable eigenvalue (see sec 2.3). This makes it
impossible for inverse iteration
to find them, since we cannot guess initial conditions that lie right
on the unstable manifold.

A different approach to the problem is that to find the zeros of the
equation $f^n(x)-x=0$, where
$n$ is the period in question. The multipoint Newton's
method \cite{webbook1} is believed to be one of the
best algorithms for that purpose. It has the advantage of converging
relatively fast to the
solution, provided that sufficiently accurate initial conditions
are given. A cycle of
period $n$ is a root of the vector function $F$: $$ F(x)= F\left(
\begin{array}{c} x_1 \\
x_2 \\ \ldots \\ x_n \\ \end{array} \right) = \left( \begin{array}{c} x_1-f(x_n)
\\ x_2-f(x_1) \\
\ldots \\ x_n-f(x_{n-1}) \\ \end{array} \right) $$ The iteration we need in order
to approach the
roots has the form $$ \frac{d}{dx}F(x)(x'-x)=-F(x)\Rightarrow
x'=x-\left[\frac{d}{dx}F(x)\right]^{-1}
F(x) $$ Therefore, after a certain number of iterations we expect to find one
root. Of course the
method has to be performed every time with different initial conditions, until
we are sure all the
solutions have been found. The issue concerning the choice of initial conditions
is a delicate one,
because in principle it is difficult to guess the initial points on a
2-dimensional plane
accurately. Nevertheless, it is useful to remember that we are dealing with
weakly coupled maps ($a\ll1$). For that reason we expect the periodic 
points to be relatively close to those
of the uncoupled system ($a=0$), that can be found analytically (see sec. 2.2) and hence exactly. 
\begin{figure}
\begin{picture} (200,200)
\put (0,150) {\rotatebox{270} {\scalebox {0.28} {\includegraphics {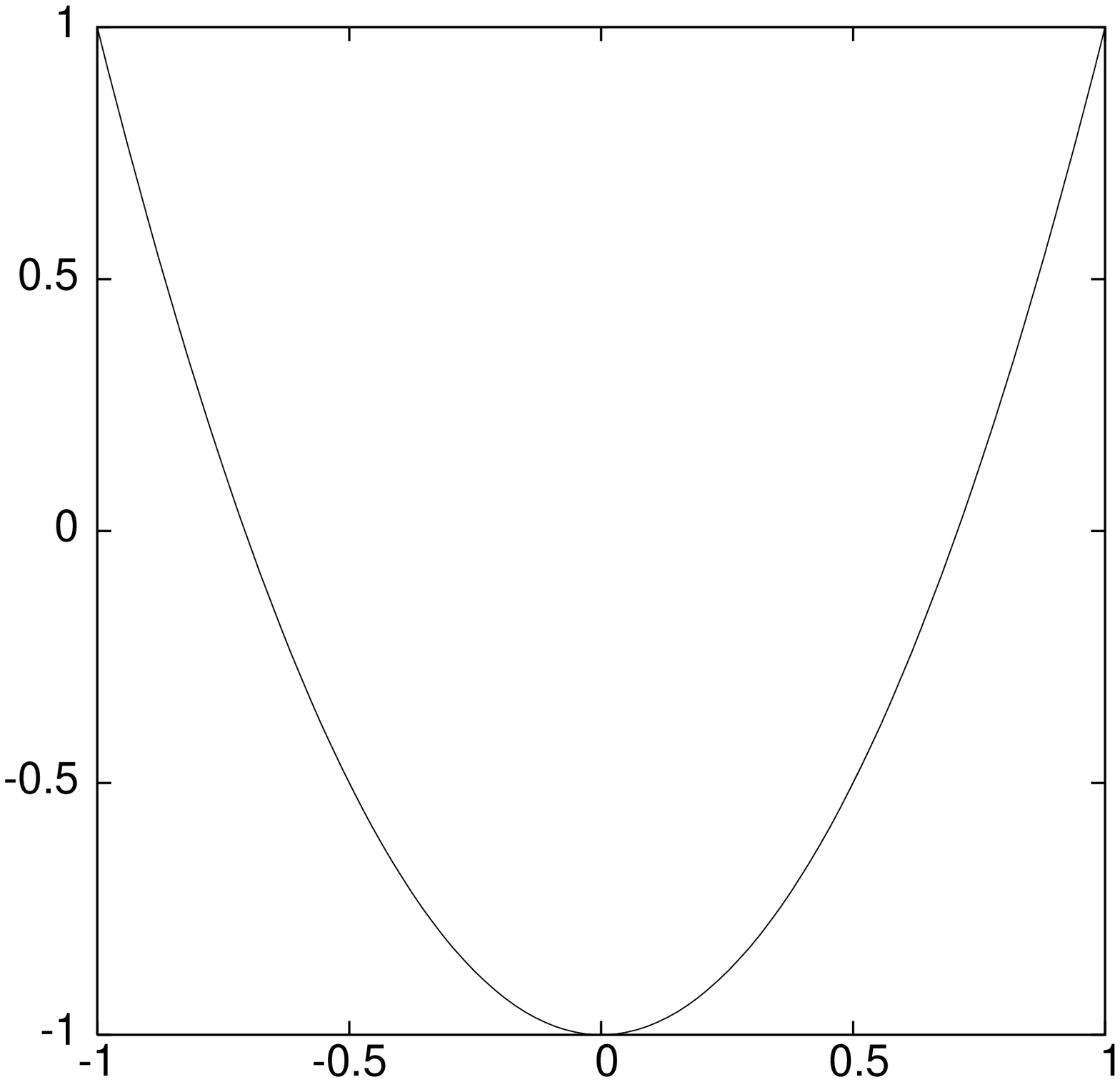}}}}
\put (180,150) {\rotatebox{270} {\scalebox {0.28} {\includegraphics {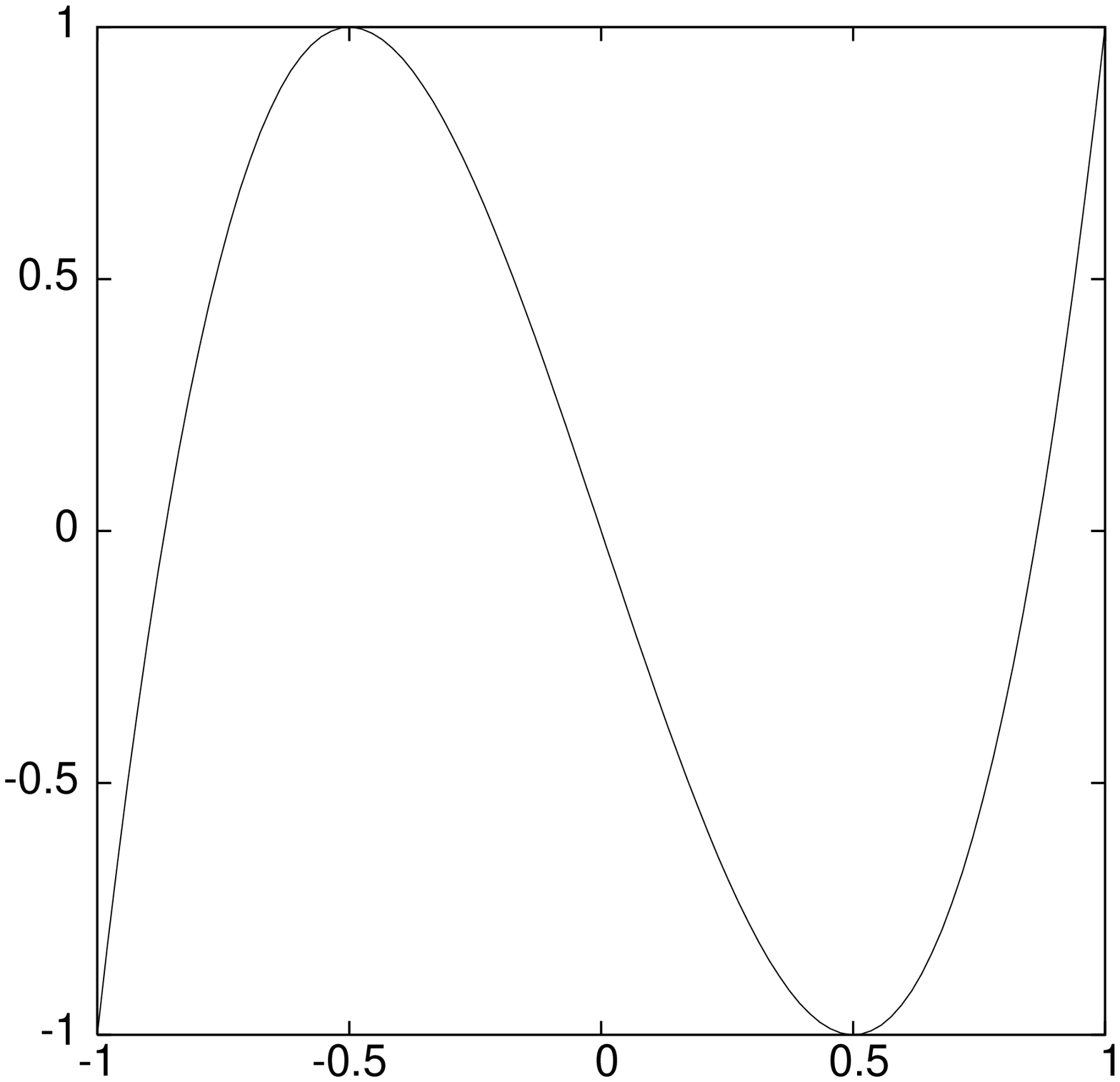}}}}
\end{picture}
\caption{$T_2$ and $T_3$ polynomials as defined in~(\protect\ref{T_N}).}
\end{figure}
\subsection{Uncoupled system}
The spatially uncoupled dynamics of Tchebyscheff maps is given by \begin{equation} \label{basic} \Phi_{n+1}^i = 
T_N(\Phi_n^i) \end{equation} where
$T_N$ is  the $N$th Tchebyscheff polynomial, defined in general as \begin{equation} \label{T_N}  T_N(\Phi)=
\cos{(N\arccos{\Phi})} \end{equation} of which we consider the first two,
namely $$
T_2(\Phi) = 2\Phi^2-1 $$ $$ T_3(\Phi) = 4\Phi^3-3\Phi $$
 The general (coupled) Tchebyscheff maps have the form \begin{equation}
\label{main} \Phi_{n+1}^i=(1-a)T_N(\Phi_n^i)+s\frac{a}{2}(T_N^b(\Phi^{i-1}_n)+T_N^b(\Phi^{i+1}_n))
\end{equation}
where $a$ is the string coupling parameter, $s$ can be $\pm 1$ and
$b$ can take on the values 1 and 0, determining 
respectively advanced coupling ($T_N^1(\Phi)= T_N(\Phi)$, ``A'') or backward
coupling ($T_N^0(\Phi)= \Phi$, ``B'').  We have then six models
$2A^+$, $2A^-$, $2B^+$, $2B^-$, $3A^+$ and $3B^+$.
The other two $N=3$ maps are equivalent to these by a trivial symmetry.
 
The model~(\ref{main}), becomes \begin{eqnarray} 
\label{sys1} 
x_{n+1}=(1-a)T_N(x_n)+saT_N^b(y_n) \\ \label{sys2} y_{n+1}=(1-a)T_N(y_n)+saT_N^b(x_n)
\end{eqnarray} on a two site lattice.
In this section we consider the uncoupled case $a=0$,
in which the system~(\ref{sys1},\ref{sys2})
reduces to \begin{equation} \label{uncoupled} x'=T_N(x) \hspace{1cm} y'=T_N(y) \end{equation}
This is useful as there is a close
correspondence between orbits at $a=0$ and orbits with $a>0$.
This correspondence is used to find orbits in the coupled dynamics 
using the uncoupled orbits as starting points for the multipoint Newton
algorithm.  The correspondence is not exact --- the $N^{2n_p}$ degree
polynomial equation giving the periodic condition always has this number of
(possibly degenerate) solutions, however for $a>0$ some of these solutions
are not in the range $[-1,1]$ or are complex, and hence not relevant to the
dynamics.  This is discussed further in the next section.

We begin first with some definitions. A {\em periodic point} of length
$n_p$ is a solution of $f^{n_p}(x)=x$; as discussed above, there are at most
$N^{2n_p}$ of them.
A {\em periodic orbit} is the orbit $\{f^j(x)\}, j=0,1,\ldots n_p-1$ of a
periodic point.  A {\em prime orbit} is a periodic orbit which is not
the repeat of a shorter orbit, that is, $f^j(x)=x$ only if $j$ is a multiple
of $n_p$.  $(x,y)$ is a periodic point of length $n_p$ of the uncoupled
system if and only if $x$ and $y$ are both periodic points of length $n_p$ of the
map $T_N$.  It is well known \cite{Beck2}
that the $T_N$ maps are conjugate to piecewise linear maps with complete
$N$-ary symbolic dynamics, by the transformation $x=\cos\pi\theta$.
This leads to exact expressions for all the periodic points of length $n_p$,
namely $x=\cos[2j\pi/(N^{n_p}\pm 1)]$ with $0\leq j\leq N^{n_p}/2$. From
these we enumerate all the prime orbits of the uncoupled two dimensional
dynamics, eliminating cyclic permutations which correspond to points on
the same orbit, and repeats of shorter orbits.

As an example we consider $N=n_p=2$.  Let symbol 0 (1) denote $x\in[-1,0]$ 
(respectively $x\in(0,1]$); in general there are $N$ symbols.  In the
symbolic representation
of periodic points, the spatial variable ($i$ in Eq. (1)) will increase to
the right, and the temporal variable ($n$ in Eq. (1)) will increase downwards. 
The periodic points of the one dimensional map are
\[ \sa{0}{0},\quad\sa{0}{1},\quad\sa{1}{0},\quad\sa{1}{1} \] 
leading to 16 periodic points in the two dimensional map, classified as
six prime orbits
\[ \sa{00}{01},\quad\sa{00}{11},\quad\sa{10}{11},\quad
\sa{01}{10},\quad\sa{00}{10},\quad\sa{01}{11} \]
their cyclic permutations
\[ \sa{01}{00},\quad\sa{11}{00},\quad\sa{11}{10},\quad
\sa{10}{01},\quad\sa{10}{00},\quad\sa{11}{01} \]
and four repeated period one orbits
\[ \sa{00}{00},\quad\sa{01}{01},\quad\sa{10}{10},\quad\sa{11}{11} \]
\subsection{Symmetry}
Until now we have not considered the symmetry of the system.  This is
particularly important for cycle expansions \cite{webbook1,Lauritzen},
for which a correct treatment
of symmetry significantly accelerates convergence of the expansions. An
uncoupled lattice has symmetry group $S_X\times G^X$ where $X$ is the number
of lattice sites (equal to two in this paper), $S_X$ is the symmetric group,
of order $X!$, corresponding to permutations of the lattice points, and $G$ is
the symmetry group of the dynamics. A symmetrically coupled lattice with $X>2$
has symmetry group $D_X$, the dihedral group, of order $2X$, corresponding
to spatial cyclic permutations and reflections of the lattice sites.  In
the case $X=2$ the cyclic shift and reflection are equivalent, so the
group is $C_2$, the cyclic group.  Remnants of the internal symmetry
group $G$ may or may not survive, depending on the nature of the coupling.
For $X>3$ the coupling breaks the permutation symmetry, reducing the order
of the symmetry group (ignoring internal symmetries), from $X!$ to $2X$ \cite{Robbins}.

Each symmetry operation permutes periodic points.  If the image of a periodic
point lies in the same periodic orbit, that orbit is called symmetric, and
is invariant under a cyclic subgroup of symmetry operations.  If it does not,
the new orbit is equivalent to the first under the symmetry, and shares many
of the same properties, in particular stability and averages invariant under
the symmetry.  It is also possible for a periodic point to be a fixed point
of the symmetry transformation, in which case its orbit is called a boundary
orbit.  In the presence of symmetry, the best strategy is
desymmetrisation, in which all images of a phase space point are considered
equivalent, and so may be described by a representative point on a subset of
the original phase space called the fundamental domain.  Equivalent orbits
in the full dynamics correspond to a single orbit in the desymmetrised
dynamics, and symmetric orbits in the full dynamics correspond to repeats of
shorter orbits in the desymmetrised dynamics.  Thus the enumeration of prime
orbits up to a given period in the desymmetrised dynamics contains much more
information than the enumeration up to the same period in the original
dynamics, hence the accelerated convergence of cycle expansions.

We continue with the previous example, considering the $N=2$ maps, which have
no internal symmetry.  Thus the symmetry group is the cyclic group $C_2$, with the nontrivial
element corresponding to $(x,y)\to(y,x)$.  The fundamental domain can be taken
as $x\geq y$, from which the full domain
is obtained by taking symmetry operations.  The boundary satisfies $x=y$.  Now
the prime orbits of period two in the desymmetrised dynamics are classified as
follows.  There are two asymmetric orbits of period two in the original
dynamics,
\[ \sa{00}{10},\quad\sa{10}{11} \]
one boundary orbit of original period two
\[ \sa{00}{11} \]
and three symmetric orbits of original period four
\[ \qa{00}{01}{00}{10},\quad\qa{01}{01}{10}{10},\quad\qa{10}{11}{10}{11} \]
Note that the total number (six) is the same, but the information content is
higher, since all the original period two orbits are included as
period one or two, and also some orbits of original period four.  
 
In the case $N=3$, the symmetry group of the interacting case is $C_2^2$,
since there is an internal symmetry $x\to-x$, which in the coupled map
survives as $(x,y)\to(-x,-y)$.  This is in addition to the permutation
$(x,y)\to(y,x)$.  Thus the fundamental domain can be taken as $x\geq|y|$.
Orbits may now be classified as asymmetric, symmetric under $(x,y)\to(y,x)$,
symmetric under $(x,y)\to(-x,-y)$, symmetric under $(x,y)\to(-y,-x)$,
boundary with $x=y$, boundary with $x=-y$ and (if it occurs) boundary
with $x=y=0$.
\subsection{Behaviour in the coupled case}
In this section we illustrate the behaviour of the $T_N$ map periodic solutions
when the coupling
parameter $a$ is in the range $[0,0.1]$. In particular, we describe two
phenomena, that is the
way and stability and number of cycles change with $a$.

\begin{figure}
\begin{picture} (500,500)
\put (50,500) {\rotatebox {270} {\scalebox {0.34} {\includegraphics {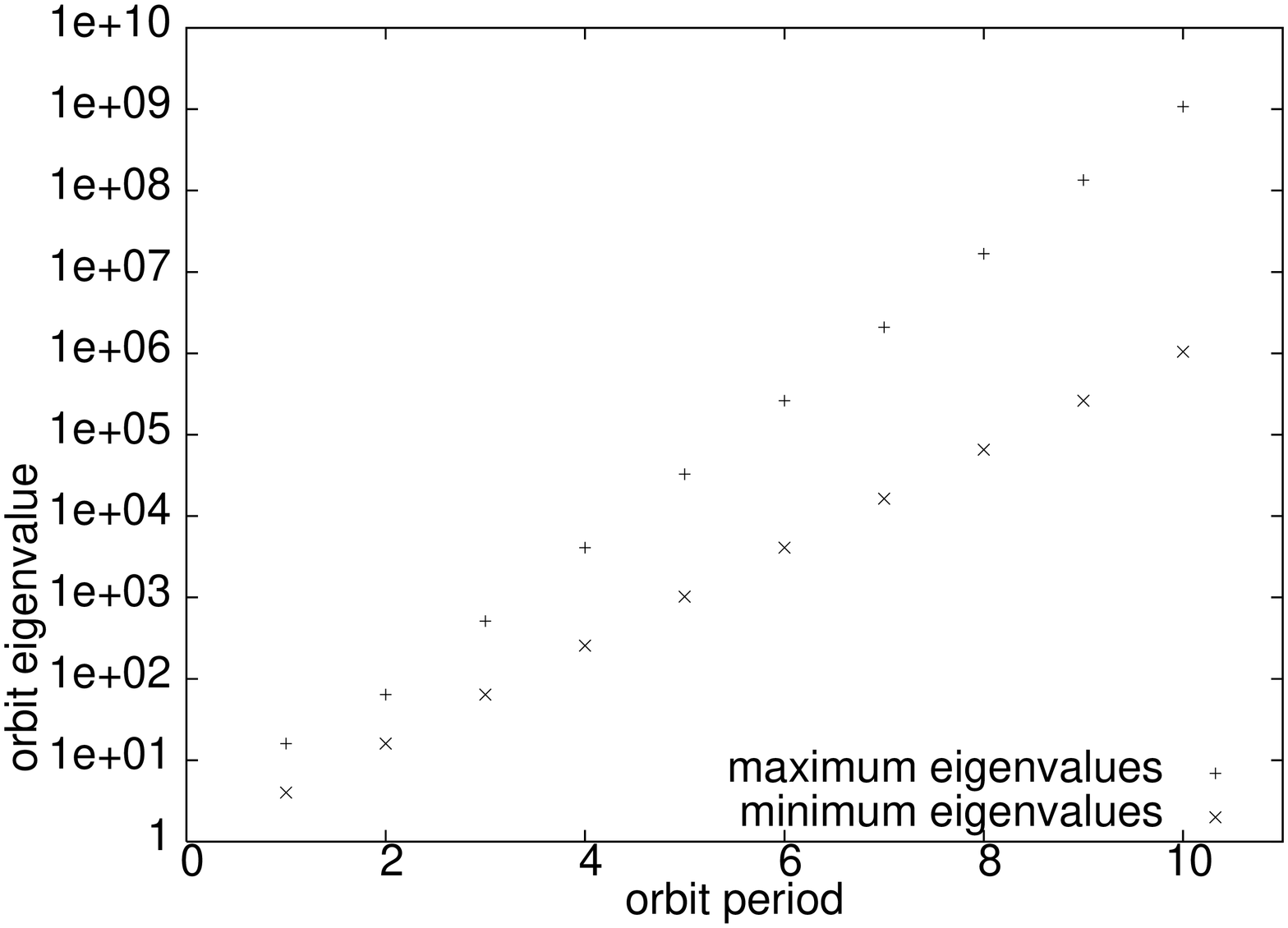}}}}
\put (50,330) {\rotatebox {270} {\scalebox {0.34} {\includegraphics {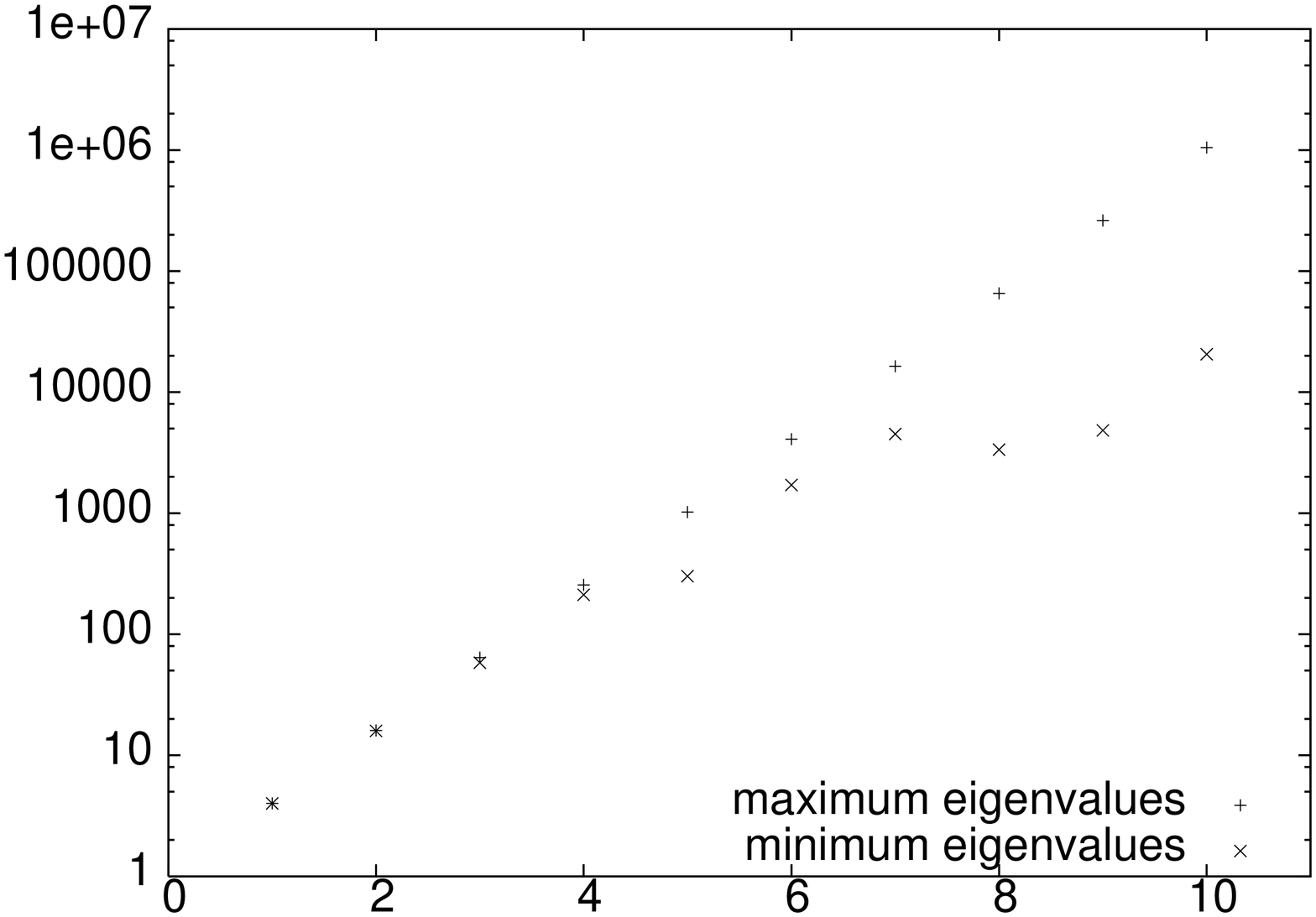}}}}
\put (50,160) {\rotatebox {270} {\scalebox {0.34} {\includegraphics {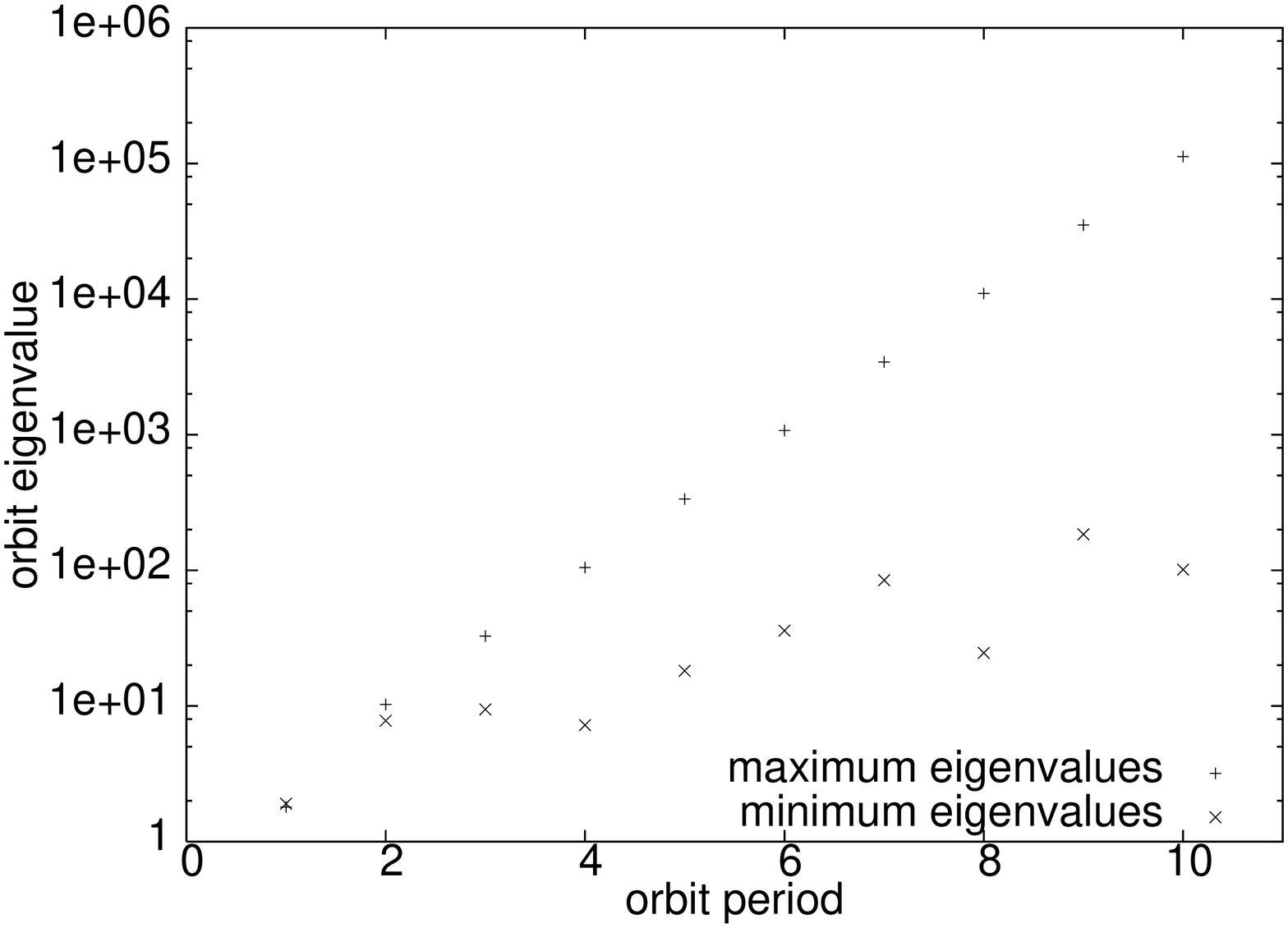}}}}
\end{picture}  
\caption {Trend of maximum and minimum eigenvalues of cycles up to period 10 of the $2A^+$ map. At the top $a=0$ (uncoupled case), in the 
middle $a=0.0001$, at the bottom $a=0.1$. 
\label{f:trend1}}
\end{figure}

\begin{figure}
\begin{picture} (150,150)
\put (50,150) {\rotatebox {270} {\scalebox {0.32} {\includegraphics 
{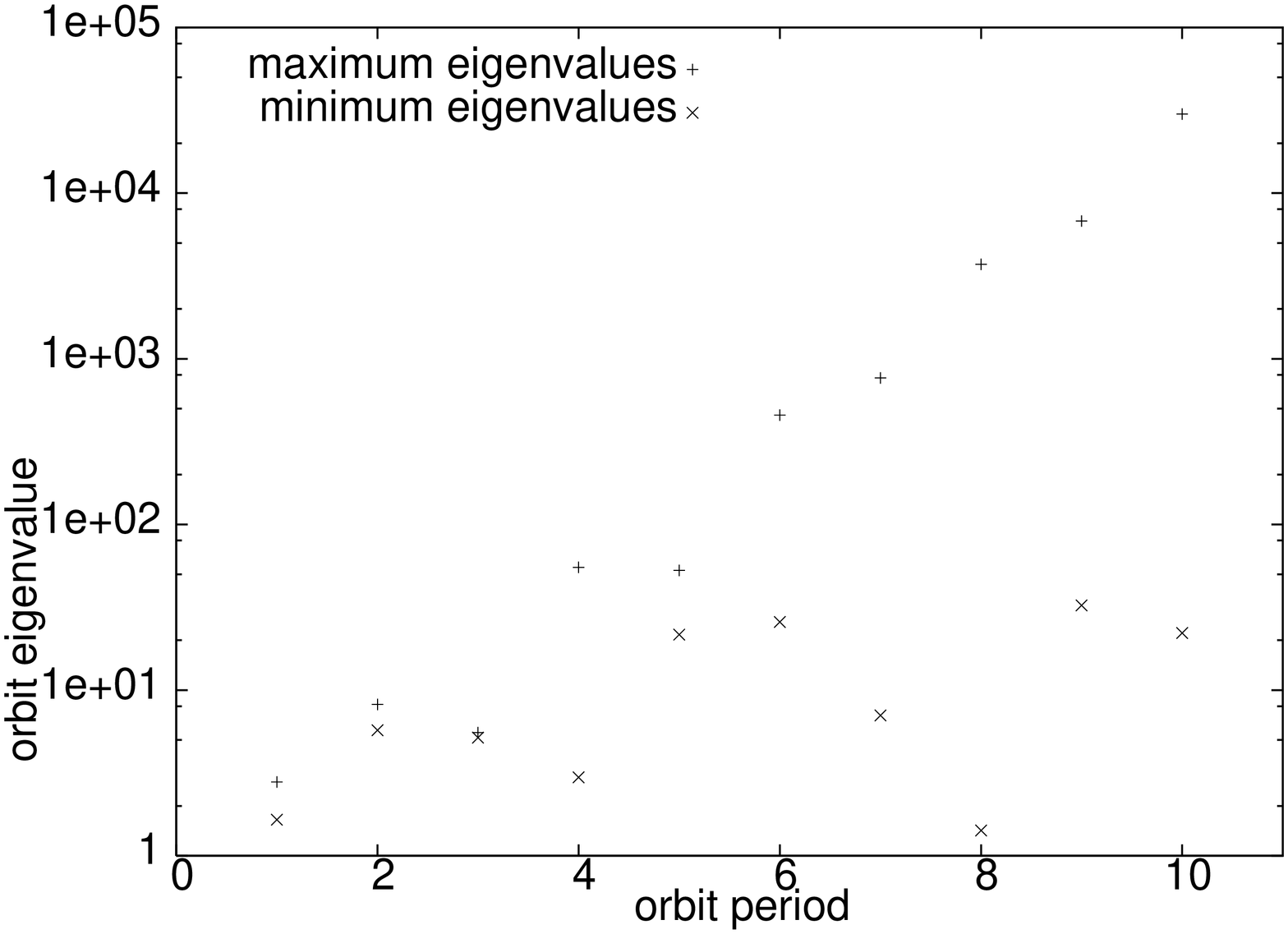}}}}
\end{picture}
\caption {maximum and minimum eigenvalues versus orbit period of the $2B^-$ map for $a=0.1$.}
\label{f:trend2}
\end{figure}

Stability is expressed by the \textit{real parts}\footnote{henceforth we call $\lambda$ eigenvalue, but we always mean its real part.}  
$\lambda_1$,$\lambda_2$ of the eigenvalues of the product of the Jacobian
matrices $$ J(x)= \left(\begin{array}{cc} \frac{\partial f(x,y)}{\partial x} &
\frac{\partial f(x,y)}
{\partial y} \\ \frac{\partial f(y,x)}{\partial x} & \frac{\partial
f(y,x)}{\partial y} \\
\end{array}\right) $$ of each point of one cycle, where $f(x,y)$ and $f(y,x)$ are the
set of equation~(\ref{sys1},\ref{sys2}). $\lambda$ is said to be stable when $|\lambda|<1$. When $a=0$, the
systems is uncoupled
and the $T_N$ maps have no stable cycles. Since the eigenvalues are continuous
functions of $a$,
we expect no stable eigenvalue to show up for very small $a$ at fixed orbit
period.   
As for stability, we analyse the
trend of the maximum and minimum eigenvalues found for each period. In general, we consider
the product $$ \Lambda=\prod_{|\lambda_i|>1}\lambda_i $$ so that in the 2-d case it
is either $\lambda_1\lambda_2$ or just the unstable $\lambda$. Stable
orbits are not used for cycle expansions, hence they are not considered for this analysis. The graphs in Figure~\ref{f:trend1} show that
both the maximum
and the minimum eigenvalues increase hyperbolically (i.e. exponentially) with respect to the period
$n_p$ of the cycles
when $a=0$ and still up to $a\sim 10^{-7}$, but the trend of the minimum
$\lambda$'s changes as
long as $a$ grows, so that their increase gets slower and slower. Besides, as we will explain later in this section, the few cycles with 
$|\Lambda|=4^{2n_p}$ move off the fundamental domain when $a\neq 0$, so that the maximum eigenvalues take the form $|\Lambda|\leq 
4^{n_p}$.\footnote{The same happens with the $T_3$ coupled maps where the maximum $|\Lambda|$ is $9^{2n_p}$ when $a=0$ and $9^{n_p}$ or lower 
otherwise.} 
When $a=0.1$, the minimum
eigenvalues are perhaps best fitted by a power law, rather than an
exponential. Such a behaviour
is common to all the coupled $T_2$ and $T_3$ maps. Figure~\ref{f:trend1} also shows that the
trend of the minimum
$\lambda$'s also gets less and less smooth, as $a$ increases. This phenomenon,
together with the
loss of hyperbolicity\footnote{for a hyperbolic system $|\Lambda|>Ae^{bn_p}$ for some $A,b>0$}, dramatically slows down the convergence of cycle
expansions, as we discuss
in the next section.
 Figure~\ref{f:trend2}  shows the trend of the eigenvalues in the $2B^-$
map when $a=0.1$.
As we can see, it is very difficult to make a good fit for the trend of the
minimum $|\Lambda|$'s
in this case, because all the coupled $T_2$ (and $T_3$) maps have got almost stable
cycles for such $a$'s, including some at high period. 
 It is interesting to compare
this behaviour with what happens with intermittent $1D$ maps \cite{intermittency}.

\begin{figure}
\begin{picture} (200,200)
\put (50,200) {\rotatebox {270} {\scalebox {0.35} {\includegraphics {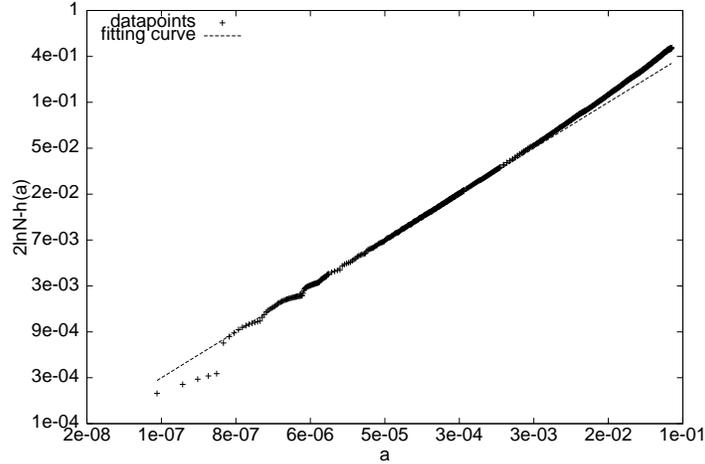}}}}
\end{picture}
\caption {$\ln [2\ln N - h(a)]$ versus $a$ in the $2A^+$ model.}
\label{f:sinking}
\end{figure}

The other interesting issue we analyse is whether and why the number of cycles
of a coupled map
for a fixed period varies as a function of the coupling parameter $a$. Numerical
simulation shows
that few periodic orbits of the uncoupled map~(\ref{uncoupled}) 
move off the fundamental domain when the system is coupled ($a>0$). Moreover, the number of $n_p\geq 4$ cycles 
continues to decrease as
$a$ grows, because some periodic
solutions of the
set~(\ref{sys1},\ref{sys2}) become complex or degenerate for certain values of $a$, so that most of the
missing orbits have
just disappeared into the complex plane. Data representing the number of periodic points $H(a,n_p)$ with $a \in 
[10^{-6},10^{-3}]$ have been fitted with the curve 
\begin{equation} \label{H}
H(a,n_p)=N^{2n_p}\exp (-2n_pCa^{\eta}) \end{equation} where  $C$ and 
$\eta$ are 
parameters that depend on the model employed, as shown in table \ref{t:tab}. 
\begin{table}
\begin{center}
\begin{tabular}{|c|c|c|} \hline model&$\eta$&$C$\\ \hline 
$2A^+$&$0.50$&$0.50$\\ 
$2A^-$&$0.50$&$0.53$\\ $2B^+$&$0.50$&$0.50$\\ $2B^-$&$0.50$&$0.54$\\ 
$3A^{\pm}$&$0.50$&$0.61$\\ $3B^{\pm}$&$0.50$&$0.63$ \\ \hline 
\end{tabular}
\caption{Parameters $\eta$ and $C$ of the fitting curve (\protect\ref{H}) with $a \in [10^{-6},10^{-3}]$, uncertainty of 
about 2 on the last digit}
\label{t:tab}
\end{center}
\end{table}   
As we can see, $\eta=0.5$ in all the models, which is theoretically plausible since the dynamics and bifurcations are 
controlled by the critical point of the Tchebyscheff map, which is quadratic. That implies that perturbations of order 
$a$ have effects of order $\sqrt{a}$, hence the exponent of $0.5$. $H(a,n_p)$ is related to the 
\textit{topological entropy} $h(a)$ of the system by $$ 
h(a)= \lim_{n_p \rightarrow \infty} \frac {\ln [H(a,n_p)]}{n_p} $$ so that $$  h(a)= 2[\ln N - Ca^{\eta}] $$ 
which decreases with a power law as $a$ increases. Figure~\ref{f:sinking} shows good agreement of the $h(a)$ fitting 
function with the data for very weak coupling ($a\lesssim 10^{-3}$)
while a slight deviation for larger $a$.  This deviation is not unexpected,
since this form for the topological entropy is only an approximate
expression in the small coupling limit.  Of course a detailed investigation
could reveal that the topological entropy is not smooth in fine detail,
however this form is remarkably accurate for practical purposes.
\section{cycle expansions and computation of averages}
\subsection{cycle expansions}
\begin{figure}
\begin{picture} (200,200)
\put(70,225) {\rotatebox {270} {\scalebox {0.38} {\includegraphics {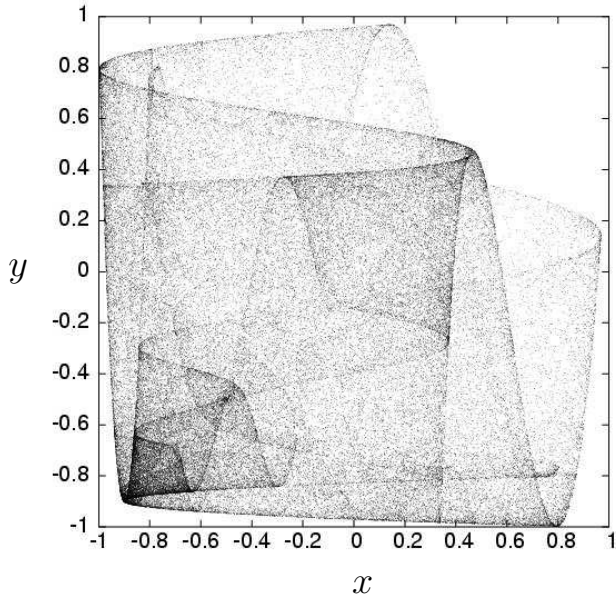}}}}
\put(60,120) {\Large $y$}
\put(190,0) {\Large $x$}
\end{picture}
\caption{Scatter plot of $(x,y)$ showing the attractor of the $2B^+$ model
at $a=0.1$.}
\label{f:attractor}
\end{figure}
\begin{figure}
\begin{picture} (220,220)
\put (0,250) {\rotatebox {270} {\scalebox {0.23} {\includegraphics {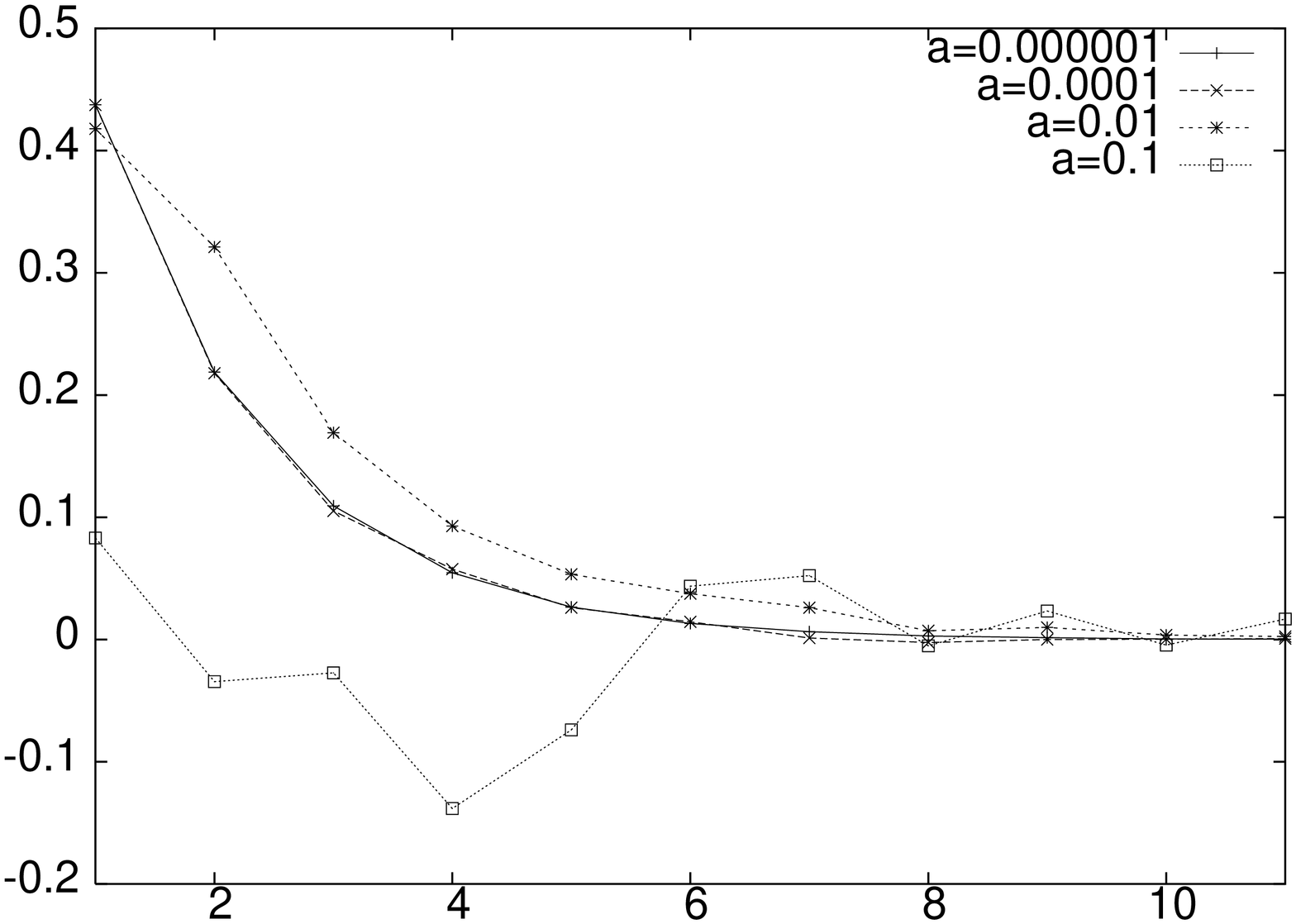}}}}
\put (170,250) {\rotatebox {270} {\scalebox {0.23} {\includegraphics {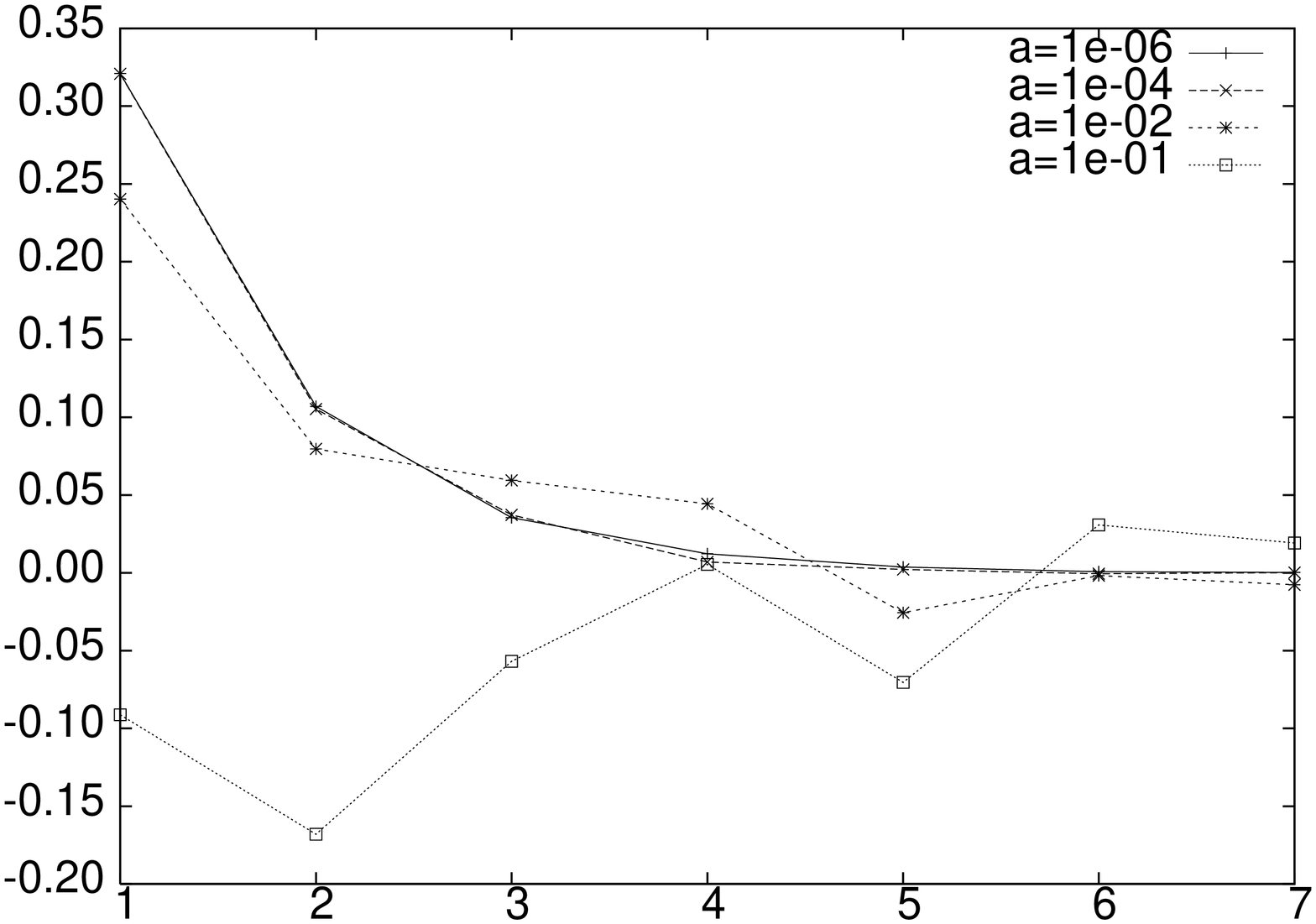}}}}
\put (0,120) {\rotatebox {270} {\scalebox {0.23} {\includegraphics {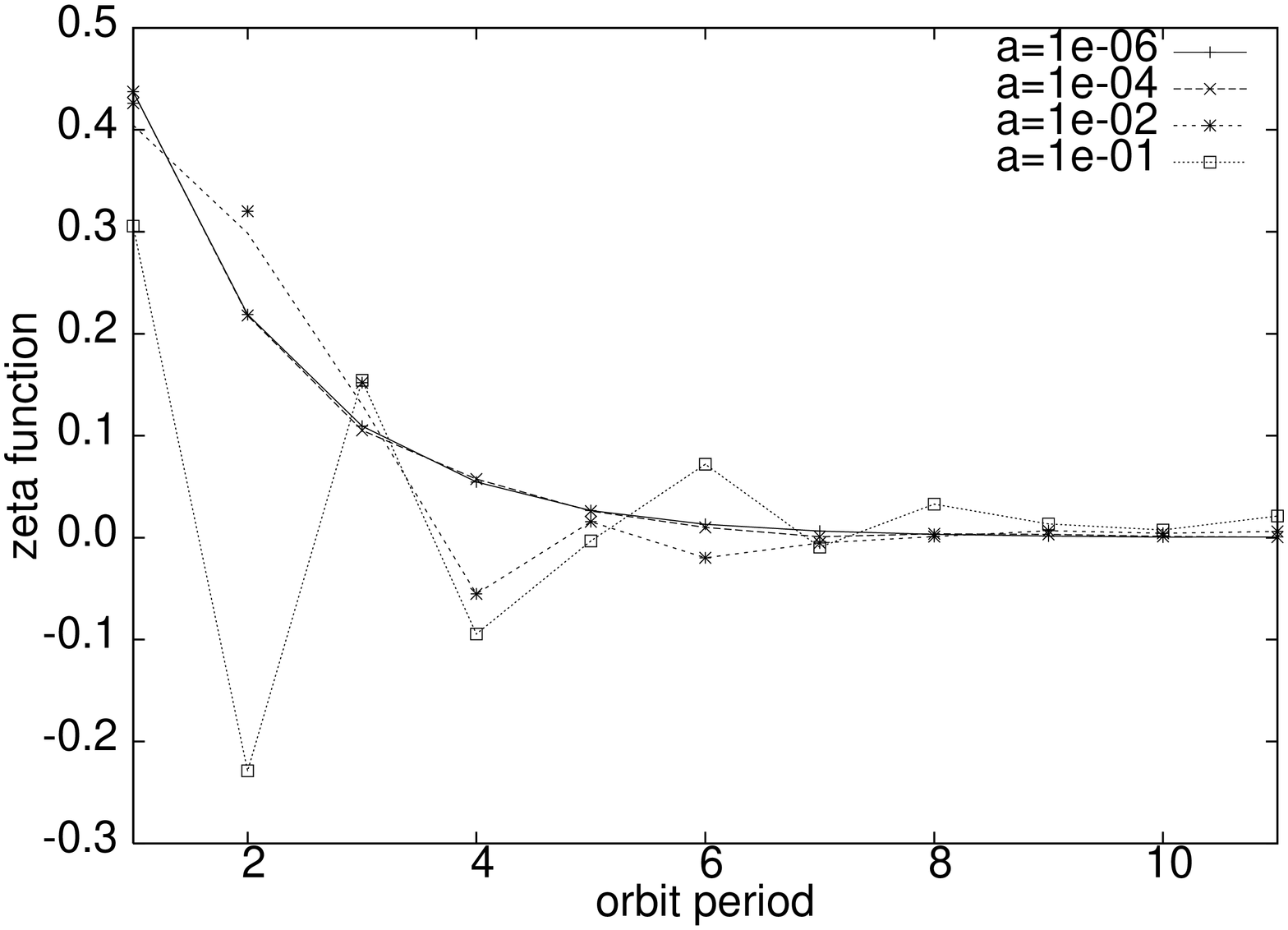}}}}
\put (170,120) {\rotatebox {270} {\scalebox {0.23} {\includegraphics {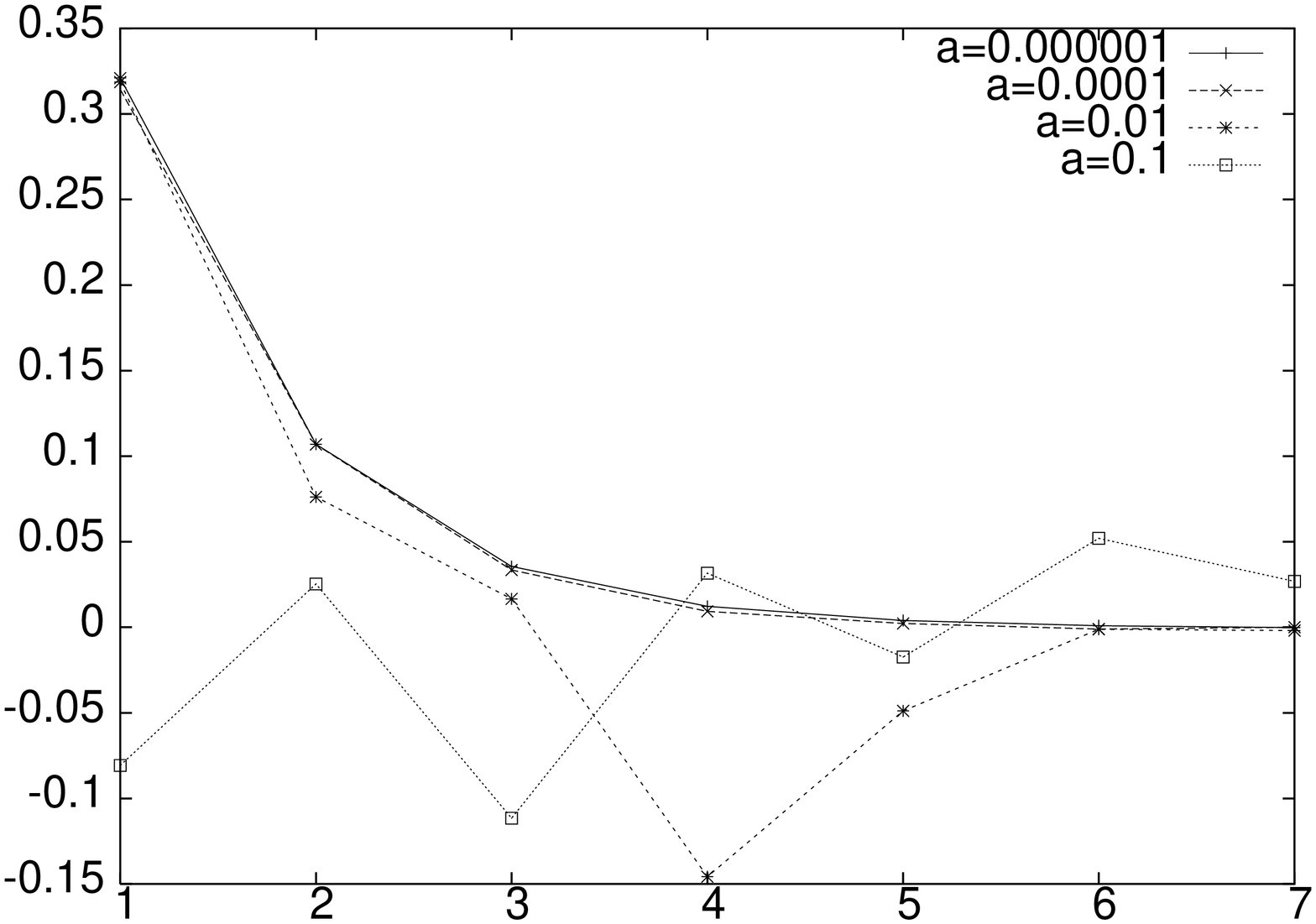}}}}
\end{picture}
\caption { Convergence of $1/\zeta(0,0)$ (logarithmic scale) versus maximum cycle length of the expansion, from the top left to the bottom right the 
maps $2A^+,3A^+,2B^+,3B^+$, shown for different values of the coupling $a$. }
\label{f:zeta}
\end{figure}

 
Given a quantity $v(x)$, we want to compute its average on the attractor of
the dynamics (\ref{sys1},\ref{sys2}), 
such as the one depicted in figure~\ref{f:attractor}.
Assuming that such attractor  
is ergodic for any value of the coupling parameter we consider,
direct simulation is the quickest method to use for this purpose,
but its precision is limited by statistics. That is one reason
why we use cycle expansions, which are potentially far more precise.
According to periodic orbit 
theory \cite{webbook1}, the average of $v$ can be found by computing the    
dynamical zeta function \begin{equation} \label{zeta} 1/\zeta = \prod_{p} \left(1-t_{p} \right) \>
, t_{p}=\frac{z^{n_p}e^{\beta V_{p}}}{|\Lambda_{p}|} \end{equation} where \begin{equation} V_{p} = \sum_{i=0}^{n-1} v(f^i(x_0)) \end{equation}     
$f$ is the map in question, $\beta$ is a parameter, $\Lambda_{p}$ the eigenvalue of the cycle $p$ and  $n$ is the orbit period. 
The uncertain digit in the average is obtained by comparing results from the computation of~(\ref{zeta}) up to the two largest periods (for example 10 
and 11).       
It is possible to check the rate of convergence of the zeta function by setting $\beta=0$ and $z=1$ and computing 
$1/\zeta(0,0)$, which should approach zero. This way we can have an idea of how effective our cycle  
expansion is and whether longer cycles are needed. Figure~\ref{f:zeta} shows the convergence 
of $1/\zeta(0,0)$ versus the truncation period of the expansion.
As we can see, the convergence is monotonic and exponential when the coupling parameter is relatively small (say $a\lesssim10^{-4}$), while it 
becomes nonmonotonic and increasingly slower for larger values of $a$.

The accuracy of $1/\zeta$ is particularly affected  towards $a=0.1$, where longer cycles or stability ordering may be needed. Ordering by stability 
means to expand the zeta function keeping all terms whose period and stability are less than a given cutoff ($|\Lambda|<\Lambda_{max}$). This scheme 
is usually preferred to length ordering in case of intermittency or bad
symbolic dynamics grammar \cite{webbook1}, but it hasn't been 
attempted here since it would need much
a higher period of truncation for the cycle expansion and, as a consequence,
impractical CPU times. A full resolution would require understanding which
orbits are close to stability, which is beyond the scope of this paper.

Anyway the behaviour shown in figure~\ref{f:zeta}  traces back to the
trend of the minimum eigenvalues of the cycles (sec. 2.3), since we can
see from equation (\ref{zeta}) 
that the smaller $|\Lambda|$'s are more relevant in the
computation of the zeta function.
That means that if their trend is exponential with respect to increasing
$a$'s, the $t_p$'s are smaller and smaller as $1/\zeta(0,0)$ approaches zero. 
Otherwise the convergence gets more complicated. The expansion for the
$T_3$ maps has been truncated at period $7$, yet the precision obtained is 
comparable with what we have got in the $T_2$ maps. That is because
$9^7\approx 4^{11}$, meaning a similar number of periodic points and hence  
more or less the same number of terms in the two expansions.   

\subsection{Computation of averages}
We now want to use cycle expansions to evaluate the average of
an observable over the two-site lattice~(\ref{sys1},\ref{sys2}) and
compare the outcomes 
with what we get with direct simulation on the same lattice. According to the theory of Beck \cite{Beck}, an interaction potential \begin{equation} 
\label{int} 
W(a)=\frac{1}{2}\Phi^i\Phi^{i+1} \end{equation} and self-interacting potentials \begin{eqnarray} \label{self2} V^{(N=2)}(a) = -\frac{2}{3}\Phi^3 + 
\Phi \\ 
\label{self3} V^{(N=3)}(a) = -\Phi^4 + \frac{3}{2}\Phi^2  \end{eqnarray} can be defined for the chaotic strings described by equation~(\ref{main}).
Zeros of~(\ref{int}) and minima of~(\ref{self2},\ref{self3}) may represent parameters of the Standard Model of particle physics such as coupling 
constants and masses of elementary particles. Since both the potentials
defined above depend on $a$ in a nontrivial way, we can only find zeros and 
minima 
numerically by averaging both $V(a)$ and $W(a)$ over the lattice~(\ref{main}). Beck has obtained his results by running a long trajectory on a 
lattice with many sites, we are going to see what happens instead by computing the zeta function~(\ref{zeta}) on the two-site 
lattice~(\ref{sys1},\ref{sys2}).   
\begin{figure}
\begin{picture} (220,220)
\put (0,250) {\rotatebox {270} {\scalebox {0.23} {\includegraphics {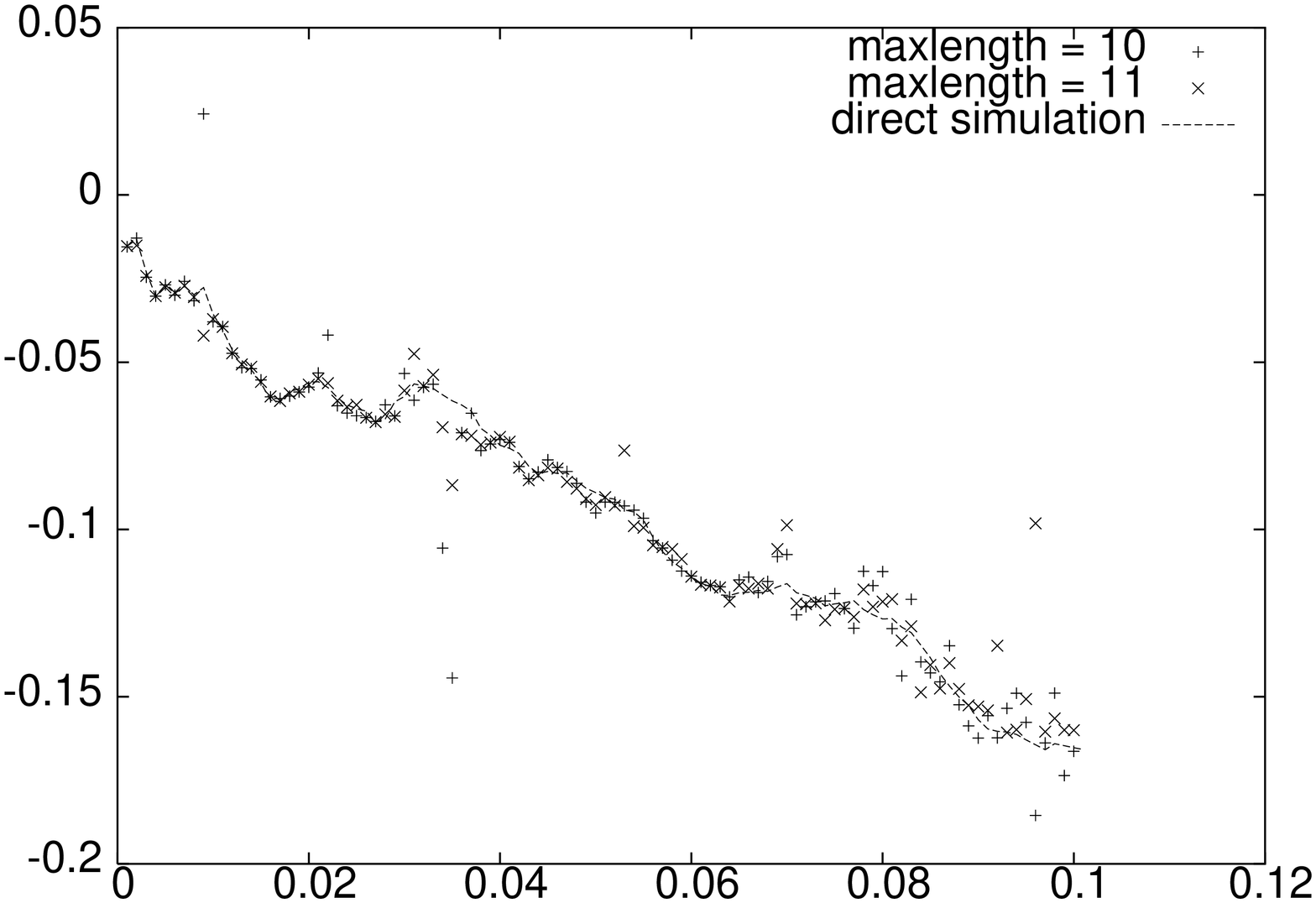}}}}
\put (170,250) {\rotatebox {270} {\scalebox {0.23} {\includegraphics {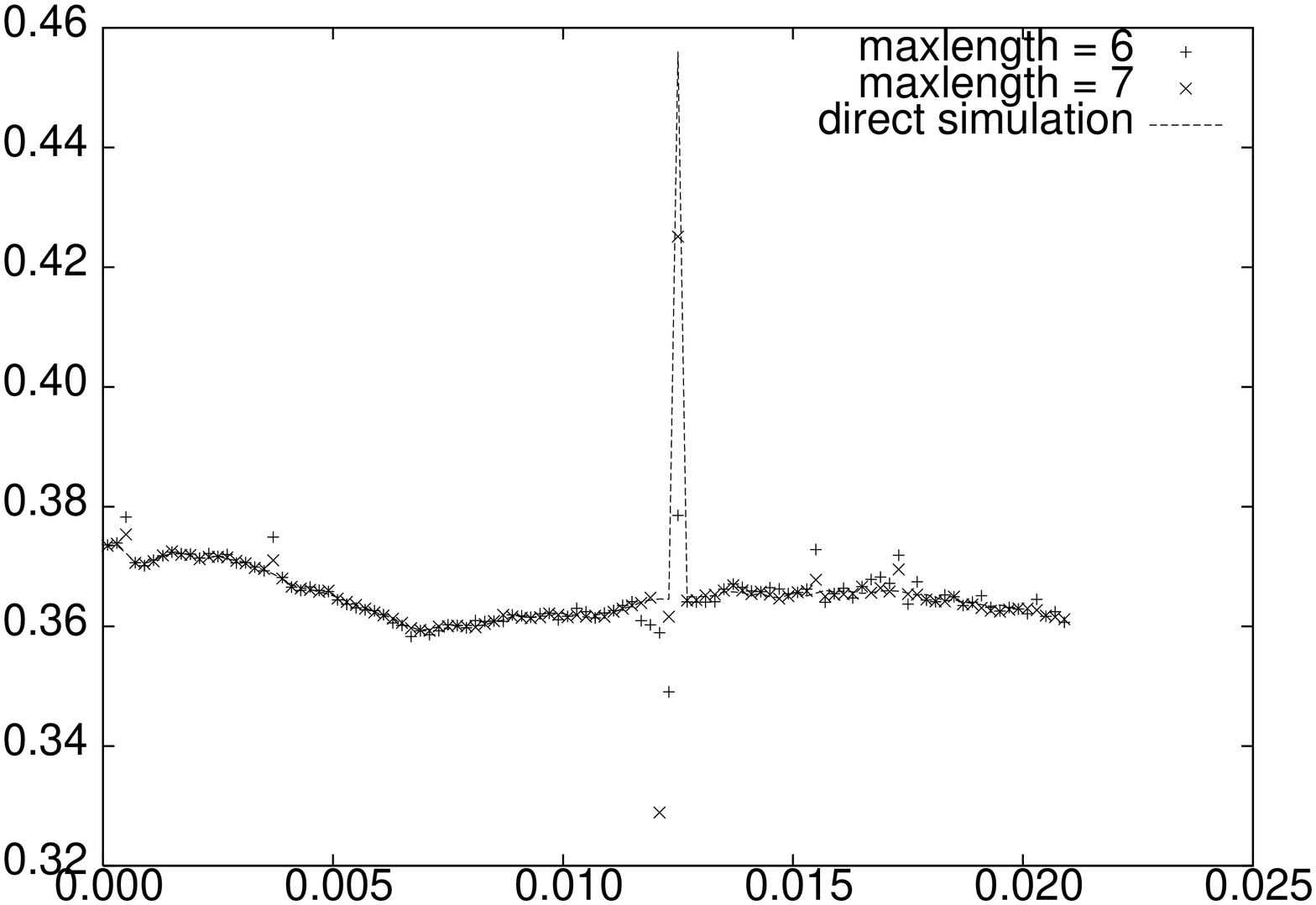}}}}
\put (0,120) {\rotatebox {270}  {\scalebox {0.23} {\includegraphics {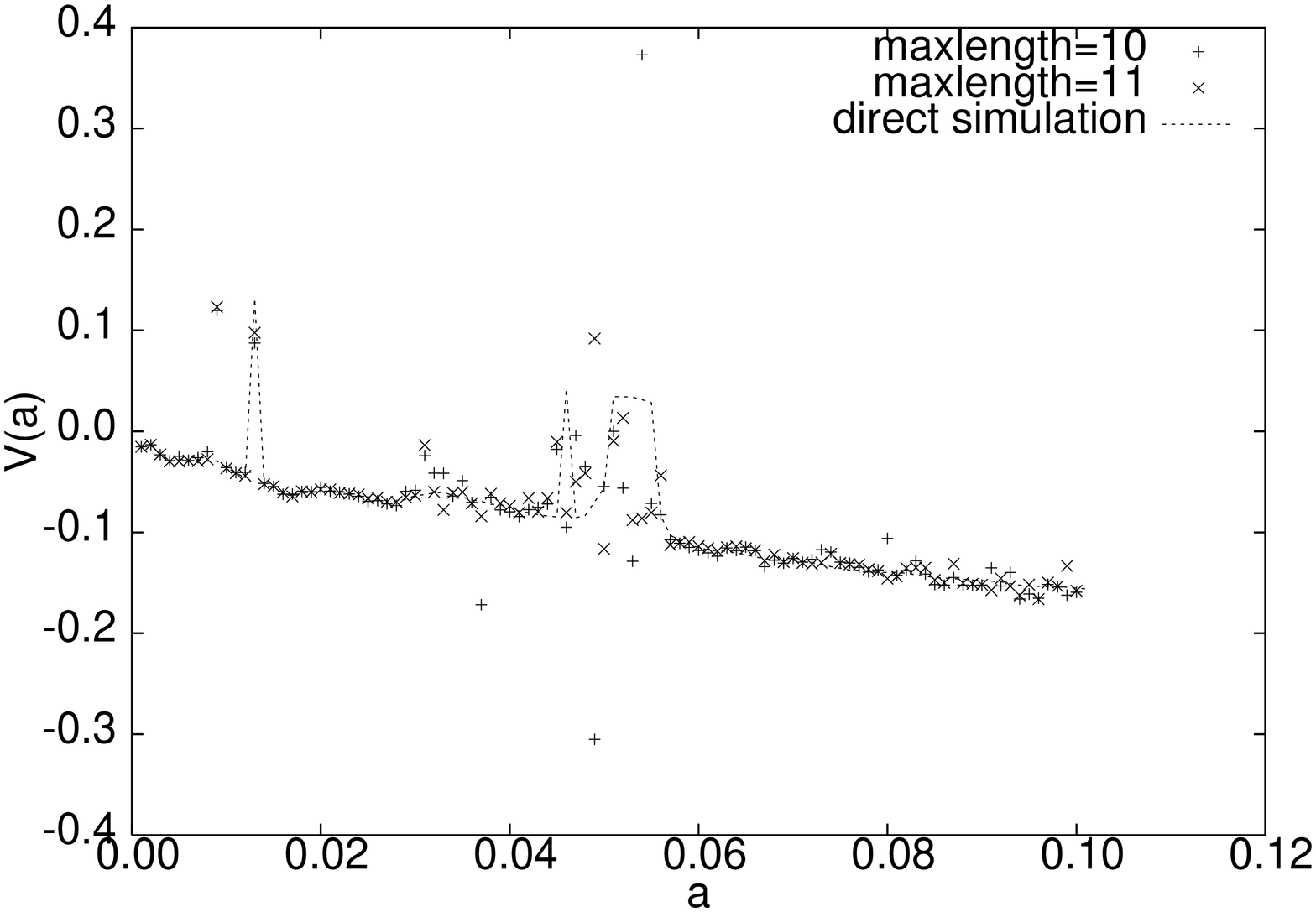}}}}
\put (170,120) {\rotatebox {270} {\scalebox {0.23} {\includegraphics {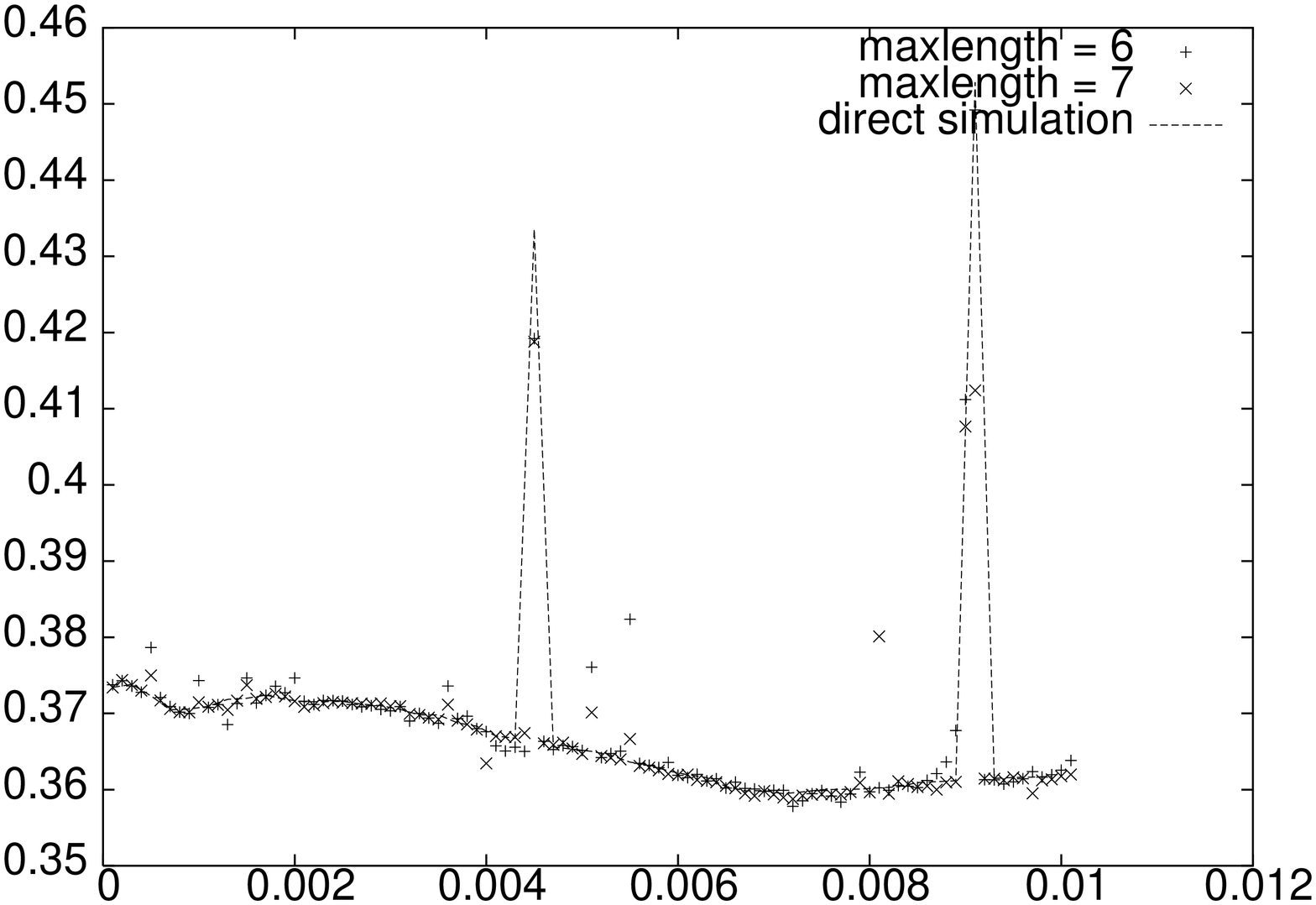}}}}
\end{picture}
\caption{ Average of the self-interacting potentials $V^{(2)}(a)$ for the $2A^+$ and $2B^+$ maps (left-hand side) and 
$V^{(3)}(a)$ for the $3A^+$ and $3B^+$ 
maps (right-hand side) versus coupling parameter $a$ , using the expansion~(\ref{zeta}) and direct simulation.}
\label{f:averages}
\end{figure}
Figure~\ref{f:averages} shows our results (achieved by computing~(\ref{zeta}) up to two different orbit periods) compared with direct simulation of 
the 
potentials $V^{(N)}$.A similar behaviour has been observed with the interaction potential $W(a)$. Five main features emerge. \begin{enumerate} \item 
In most points cycle expansion coincides with direct simulation. 
\item Cycle expansion is far offset on some 
points, which can be explained with the presence of almost stable cycles ($|\Lambda|\sim 1$) of high periods whose terms in eq.~(\ref{zeta}) have too 
much weight in the expansion. \item There are points where both cycle
expansion and direct simulation are offset. That is probably caused by
one or more stable 
cycles or chaotic attractors, whose basins of attraction alter the phase
space making it no longer ergodic. \item The precision of the expansion 
visibly decreases as $a$ 
approaches $0.1$, as we expect from the observations made in the previous sections on the trend of the minimum eigenvalues and the behaviour of 
$1/\zeta(0,0)$. \item Figure~\ref{f:screwed} shows an offset at some point between cycle expansion and direct simulation in the model $2B^+$ performed 
with different starting values from those used in figure~\ref{f:averages}. That proves a  
lack of ergodicity in some regions of the interval $[0,0.1]$ of $a$ and indicates that the unstable periodic
orbits are mostly unaffected by the presence of a stable
orbit (or other ergodic component such as an elliptic region).  
Hence the cycles  form a transient chaotic set more or less covering
the attractor (Fig.~\ref{f:attractor}) for similar values of the parameter.
In this sense, the periodic orbit expansion results are more stable
than direct simulation with respect to variation of the parameter,
and may be preferred for this reason. \end{enumerate}
\begin{figure}
\begin{picture} (140,140)
\put (70,170) {\rotatebox {270}  {\scalebox {0.30} {\includegraphics {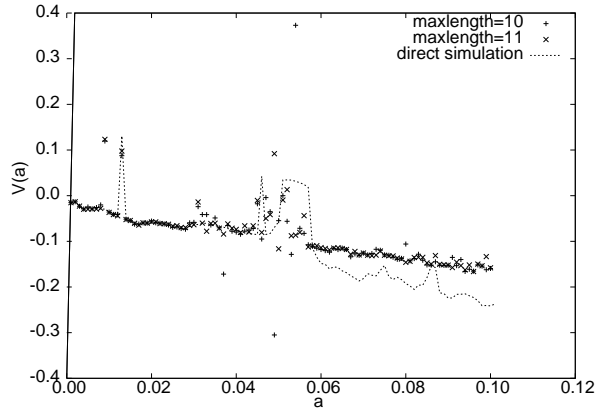}}}}
\end{picture}
\caption{ Average of the self-interacting potential $V^{(2)}(a)$ in the $2B^+$ model obtained with cycle expansion and direct simulation (with 
different 
starting values from those in figure~\ref{f:averages}) versus coupling parameter $a$.}
\label{f:screwed}
\end{figure}
\section{Conclusions and outlook}
Periodic orbits have provided a unique window with which to view properties
of weakly coupled CML.  The periodic orbits in the uncoupled limit provide
a reliable means of finding periodic orbits of the weakly coupled system.
The symmetry of the coupled system has been exploited to improve convergence
of the cycle expansion.  Cycle expansion techniques work as expected, well
near the uncoupled, more chaotic limit, but with slower convergence
at stronger coupling.  The computation of averages using cycle expansions
leads to important information at a glance concerning the number and
nature of nearly or fully stable states coexisting with the chaotic web
of unstable periodic orbits.

The observed power law dependence of the topological
entropy with parameter requires a more detailed investigation of the
density and nature of bifurcations in this system to explain the
value of the power $\eta=1/2$, the extent of its apparent universality
and corrections for finite values of the parameter.
For a more rigorous approach to the topological entropy of a different
two site model, see \cite{bastien}.

This work has clarified a number of issues and challenges regarding a
more general
periodic orbit approach to CML.  For stronger coupling (hence weaker
chaos) stability ordering of cycle expansions \cite{webbook1} is
likely to be more
effective than the length ordering employed here; this requires
an understanding of the symbolic dynamics of the most stable cycles,
which may be obtained in the study of the bifurcations mentioned above.
Larger lattices are of course an important aspect, however the rapid
decay of spatial correlations for weakly coupled CML make this less
important than it might seem.  The fact, as observed here, that small
windows of stability at weak coupling affect the cycle expansions at
only a few parameter values, gives great confidence since their presence
is expected to be even less evident in lattices of larger size, in which
typical orbits have many unstable directions.
\section{Acknowledgements} 
We acknowledge financial support of the Nuffield Foundation
(grant NAL/00353/G) and helpful discussions with P. Cvitanovi\'c.
DL would also like to thank G. Servizi and
V. Smith for having made available computing facilities in Bologna and
Bristol respectively,
as well as N. Chevalier, R. Morris and A. Shall for technical support.

 \end{document}